\documentclass[12pt]{article}
\usepackage{amsmath, amsthm, amssymb}
\usepackage[dvips]{graphics}
\usepackage{xcolor}
\usepackage{subfigure}
\usepackage{multirow}
\usepackage{graphicx} % needed for graphics
\usepackage{bm}     % provides bold greek letters in math mode
\usepackage{setspace} % this gives me the ability to double space
\usepackage{booktabs} % gives better spacing in tables
\usepackage{verbatim}
\usepackage{appendix} % this gives an appendix title to the section
\usepackage{fullpage} % automates margin size etc.
\usepackage{natbib} % for bibliography
\usepackage{changepage}
\usepackage{caption}
\usepackage{geometry}
\usepackage{hyperref}
\usepackage{xr-hyper}
\usepackage[bb=boondox,scr=rsfso]{mathalfa} % replace its Blackboard bold and use rsfso for \mathscr.

\usepackage{soul}

\theoremstyle{plain}

\theoremstyle{definition}

\theoremstyle{remark}

\newcommand{\bx}{\bm{x}}
\newcommand{\by}{\bm{y}}
\newcommand{\bz}{\bm{\xi}}

\newcommand{\bxs}{\bx^\star}
\newcommand{\bys}{\by^\star}
\newcommand{\bxo}{\bx^o}
\newcommand{\bxso}{\bx^{\star o}}
\newcommand{\bths}{\bm{\theta}^\star}
\newcommand{\bxi}{\bm{\xi}}
\newcommand{\bzeta}{\bm{\zeta}}

\newcommand{\mus}{\mu^\star}
\newcommand{\sig}{\sigma}
\newcommand{\sigs}{\sig^{2\star}}

\newcommand{\tg}{\tilde{g}}

\newcommand{\OO}{\mathcal{O}}

\newcommand{\CC}{\mathcal{C}}

\newcommand{\Cjl}{\mathcal{C}_{j\ell}}
\newcommand{\Ctjl}{\tilde{\mathcal{C}}_{j\ell}}
\newcommand{\xjlo}{\bm{x}_{j\ell}^{\star o}}
\newcommand{\defeq}{\stackrel{\mathrm{def}}{=}}

\bibpunct{(}{)}{;}{a}{}{,}

\makeatletter
\def\moverlay{\mathpalette\mov@rlay}
\def\mov@rlay#1#2{\leavevmode\vtop{%
   \baselineskip\z@skip \lineskiplimit-\maxdimen
   \ialign{\hfil$\m@th#1##$\hfil\cr#2\crcr}}}
\newcommand{\charfusion}[3][\mathord]{
    #1{\ifx#1\mathop\vphantom{#2}\fi
        \mathpalette\mov@rlay{#2\cr#3}
      }
    \ifx#1\mathop\expandafter\displaylimits\fi}
\makeatother

\newcommand{\iid}{\stackrel{iid}{\sim}}
\newcommand{\ind}{\stackrel{ind}{\sim}}

\begin{document}
\onehalfspace

%\title{Prediction in the Presence of Missing Covariates}
\title{ Clustering and Prediction with Variable Dimension Covariates}
\author{ Garritt L. Page \\ Department of Statistics \\ Brigham Young University, Provo, Utah   \\ BCAM - Basque Center for  Applied Mathematics, Bilbao, Spain \\page@stat.byu.edu
         \and	Fernando A. Quintana \\ Departamento de Estad\'{i}stica \\ Pontificia Universidad Cat\'{o}lica de Chile, Santiago
         \\ and Millennium Nucleus Center for the \\ Discovery of Structures in Complex Data \\ quintana@mat.uc.cl
         \and Peter M\"uller \\ Department of Mathematics \\ The University of Texas at Austin \\ pmueller@math.utexas.edu
}
\maketitle
\begin{abstract}
In many applied fields incomplete covariate vectors are commonly
encountered. It is well known that this can be problematic when making
inference on model parameters, but its impact on prediction performance is
less understood. We develop a method based on covariate dependent
partition models that  seamlessly handles missing covariates while
completely avoiding any type of imputation. The method we develop  allows
in-sample predictions as well as out-of-sample prediction,  even if the
missing pattern in the new subjects' incomplete covariate vector was not
seen in the training data.  Any data type, including  categorical or
continuous covariates are permitted. In simulation studies the proposed
method compares favorably.  We illustrate the method  in two application examples.
\end{abstract}

{{\bf Key Words}: Dependent random partition models, indicator-missing,
pattern missing, Bayesian nonparametrics}

\doublespace

\section{Introduction}
We introduce an approach for prediction with missing covariates, that is,
regression with a variable-dimension covariate vector. The proposed model
does not require any notion of imputing or substituting
  % or other assumptions about the
missing covariates. Instead we start with a  distribution for a  random
partition based on available covariates, and then add a cluster-specific
sampling model for the response. The result is an elegant and
uncomplicated variable-dimension regression approach.

Missing observations are regularly encountered in data-driven research
(\citealt{daniels&hogan:2008}, \citealt{molenberghs2014handbook}). Because
of this, there is a rich literature dedicated to methods that have been
developed to accommodate them.  These methods range from being ad-hoc like
the complete-case approach which simply deletes subjects/units  exhibiting
missing observations,  to more statistically sound procedures like
(multiple) imputation which probabilistically ``fills'' in the missing
values (see \citealt{rubin:1987}, \citealt{Little&RubinBook},
\citealt{van2012flexible}, or \citealt{molenberghs2014handbook}).  Most of
the statistical literature dedicated to missing observations is focused on
missing response values and their impact on inference for model
parameters. The focus of this work is on incomplete covariate vectors and
their impact on prediction accuracy. Even though incomplete predictor
vectors are ubiquitous and can have adverse effects on prediction accuracy
(destructive if an influential predictor is missing), the missing
observations literature is much sparser for this case.

In the presence of missing covariates the complete-case approach is still
an option, but often performs poorly when prediction is of interest
(\citealt{mercaldo&bluem:2018}). Some multiple imputation methods that
were developed for missing response values can also be employed for
missing covariates.  Focusing on methods that allow mixed data types,
multiple imputation by chained equations (MICE), which employs
conditionally specified models, can be used to impute missing covariates
one-at-a-time (\citealt{van2012flexible}). This approach is somewhat
ad-hoc as there is no guarantee that the conditionally specified models
produce a valid joint model for the covariates. To avoid this,
\cite{xu&daniels&winterstein:2015} employ Bayesian additive regression
trees (BART) to impute missing covariates based on the MICE framework.
Although their approach produces a valid joint distribution,  the order of
the conditional models impacts the imputations. Similarly,
\cite{burgette&reiter:2010} employ classification and regression trees
(CART) to impute within a MICE type algorithm which permits more
flexibility in the conditional distributions,  while
\cite{stekhoven&buhlmann:2012} use random forests to carry out imputation.
Recently, \cite{storlie&Therneau&etal:2019} built a flexible yet complex
Bayesian nonparametric model to carry out imputation. Their approach
jointly models mixed-type covariates and includes a variable selection
component making the procedure more robust. All these and most other
multiple imputation type approaches focus on inference for model
parameters. If prediction is tangentially considered, the complications
that arise when predicting based on multiple imputation are not
considered. For example, procedures based on multiple imputation are
problematic when out-of-sample prediction is desired as it is not possible
to connect a response to the vector of covariates when carrying out
imputation (a response does not exist). This has been shown to negatively
impact predictive performance
(\citealt{moons&donders&stijnen&harrell:2006}). Considering these
limitations, our interest lies in developing a procedure that avoids
% requires absolutely no type of
imputation while still providing a good and flexible model for the
available data.

The so called missing indicator approach (\citealt{little:1992},
\citealt{jones:1996}, \citealt{geert:2006}) has been developed to avoid
the sometimes unverifiable assumptions of multiple imputation. But these
methods must be used with care in practice as they are prone to producing
biased estimates, and as a result, poor predictions (see
\citealt{geert:2006} and \citealt{Groenwold:2012}).  Also, under this
approach, there is no clear way to handle the case of a new subject in
out-of-sample prediction exhibiting a different missing pattern than those
found in the training data.

Our approach to incorporating missing covariates in a prediction model
stems from  a completely different perspective. We start with  a
covariate-dependent random partition model that naturally allows for
missing values in the covariates, and can accommodate  mixed-type
covariates. Adding a cluster-specific sampling model to the random
partition defines a posterior predictive distribution that makes
out-of-sample prediction straightforward. Covariate-dependent random
partitions are particularly well suited for prediction, as they permit
complex interactions and nonlinear associations between covariates and
responses, simply by including corresponding clusters in the partition.
Perhaps the  missing data method whose focus is most similar to what we
develop is found in \cite{kapelner&bleich:2015}. They use BART to
 implement  predictions and employ a missing indicator when
constructing trees (i.e., the trees are not used as a tool to
impute). Although their motivation is similar to ours, our approach is
based on partitions, which permits more flexibility in capturing how
covariates interact. Throughout we assume that missing data are
missing at random (MAR), with some exceptions. Simple MNAR due to a
detection limit, for example, is easily accommodated by introducing an
additional binary covariate.
%We explore this case as part of a simulation study by including a scenario
%with MNAR in the simulation truth. \footnote{This is Peter's text sent via
%e-mail.  We didn't include an additional covariate in the simulations or
%data examples } and their associations.

The remainder of the article is organized as follows. In
Section~\ref{sect:model} we provide background associated with covariate
dependent product partition models. Section~\ref{sect:PPMxMissing}
describes  our extension that permits incomplete  covariates vectors of
varying dimensions. Section~\ref{simulation.study} contains a simulation
study while data applications are described in Section~\ref{sect:data}.
Some concluding remarks are provided in Section~\ref{sect:concl}.

\section{A Covariate-Dependent Product Partition Model}\label{sect:model}

We build on   a covariate dependent partition model proposed
by~\cite{PPMxMullerQuintanaRosner}. We  introduce notation by way of a
brief review of the approach. For more details see
\cite{PPMxMullerQuintanaRosner},  \citet{bgPPM}, or
\cite{quintana&loschi&page:2018}.

Let  $i = 1, \ldots, m$ index $m$ experimental units. Let $\rho_m = \{S_1,
\ldots, S_{k_m}\}$ denote a partition (or clustering) of the $m$ units
into $k_m$ nonempty and exhaustive subsets so that $\{1,\ldots,m\} =
\bigcup_j S_j$, for disjoint subsets $S_j$. To simplify notation we omit
the subscript $m$ for $\rho$ unless explicitly needed. A common
alternative representation of $\rho$ introduces cluster membership
indicators $c_i = j$ if $i \in S_j$. Let $\bx_i = (x_{i1}, \ldots,
x_{ip})$ denote a $1\times p$ covariate vector measured on unit $i$ and
$\bx = \{\bx_1, \ldots, \bx_m\}$.
  % will denote a matrix of ``stacked'' covariate vectors.
Further, let $\bx^{\star}_j = \{\bx_i:i \in S_j\}$ denote covariate
vectors arranged by clusters. We will generally use a superscript
``$\star$'' to mark cluster-specific entities. The covariate-dependent
product partition model (PPMx) prior on $\rho$  formalizes  the idea that
units with similar covariate values are more likely {\it a priori} to
belong to the same cluster than  units with dissimilar covariate values.
The prior consists of two set functions. The first, called a cohesion
function and denoted by $c(S_j\mid M) \ge 0$ for $S_j \subset \{1, \ldots,
m\}$ and  hyper-parameter $M$, measures prior belief associated with the
co-clustering of the elements of $S_j$. The second, called a similarity
function and denoted by $g(\bx^{\star}_j\mid\bxi)$ and parametrized by
$\bxi$, formalizes the ``closeness'' of the $x_i$'s in a cluster by
producing larger values of $g(\bx^{\star}_j\mid\bxi)$ for $x_i$'s that are
more similar.  The similarity function in the PPMx plays a similar role to
that of the impurity function when building trees using CART
(Classification And Regression Trees). See, for example,  Section~2.4
in \cite{sutton:2005}.  With the similarity and cohesion functions, the
form of the PPMx  prior is the following product
\begin{align} \label{ppmx}
p(\rho\mid\bx, M,\bxi) \propto \prod_{j=1}^{k_m} c(S_j\mid M)g(\bx^{\star}_j\mid\bxi).
\end{align}
The cohesion function we employ in what follows is $c(S_j\mid M) =
M\times(| S_j| - 1)!$ for some positive $M$ and $\mid\cdot\mid$ denoting
cardinality. This cohesion is commonly employed as the corresponding prior
$p(\rho)$ is identical to the popular Chinese restaurant process
\citep{broderick&jordan&pitman:13}.
% and its popularity stems from connections
% the resulting random partition model has to that which is based
% on
% a Dirichlet process model~\citep[see, e.g.,][]{quintana:06}.
Regarding possible similarity functions, \citet{PPMxMullerQuintanaRosner}
discuss choices  for different covariate data types  (continuous, ordinal,
or categorical),  and suggest using
\begin{align}\label{sf}
g(\bxs_j\mid\bxi) = \int\prod_{i \in S_j} q(\bx_i \mid\bm{\zeta}_j) q(\bm{\zeta}_j\mid\bxi) d\bm{\zeta}_j.
\end{align}
Here, $\bxi$ are fixed hyper-parameters. The integral can be evaluated in
closed form if $q(\bx_i \mid \bzeta_j)$ and $q(\bzeta_j \mid \bz)$ are
chosen as a conjugate sampling model and prior pair. The model $q(\cdot)$
is used only for easy calculus, rather than any notion of statistical
modeling (indeed $\bx$ may not even be a random variable). They are only
used as a means to measure the agreement of the covariates  in $\bxs_j$,
implicitly defining ``similar'' as high marginal probability under
$q(\cdot)$. But any other function that assigns large values for $\bxs_j$
that are judged to be similar can be used. For simplicity we use
\eqref{sf} for scalar covariates only and construct $g(\cdot)$  for
multivariate $\bx_i$ using separate similarity functions $g_\ell$  for
each covariate and set $g(\bx^{\star}_j\mid\bxi) = \prod_{\ell = 1}^p
g_\ell(\bx^{\star}_{j \ell}\mid\bxi_{\ell})$ where $\bx^{\star}_{j \ell} =
\{x_{i\ell} : i \in  S_j \}$.  See \cite{page:2018} for more discussion on
other possible specifications for the similarity function.

For a given cluster arrangement $\rho$, we complete the model construction
with a sampling model for the response $y_i$ by introducing
cluster-specific parameters $\bths = (\bths_1, \ldots, \bths_{k_m})$ and
assuming conditional independence at the observation level. Letting $y_i$
denote the $i$th response and $\bm{y} = (y_1, \ldots, y_m)$ this leads to
the following model
\begin{eqnarray}
  p(\by, \rho, \bths, M, \bxi \mid \bx) &=&
    {p(\by \mid \rho, \bths)}\, p(\rho\mid\bx,M,\bxi)p(\bths) \nonumber \\
   & \propto & \prod_{j=1}^{k_m}
               \left\{ \left( \prod_{i\in S_j}p(y_i \mid \bths_j) \right)
               p(\bths_j) \right\}
          p(\rho \mid \bx,M,\bxi),
          \label{eq:py}
\end{eqnarray}
where $p(\bths)$ is a prior distribution for $\bths$  whose components are
assumed to be independent and identically distributed. The model can be
written in hierarchical form using  latent cluster membership indicators,
\begin{align}\label{HierModel}
\begin{split}
  y_i \mid \bths, c_i=j  &  \ind p(y_i\mid \bths_{j}) \\
        \bths_j  \mid\rho & \iid p(\bths_j\mid\rho)\\
        p(\rho=\{S_1, \ldots, S_{k_m}\} \mid \bx,M, \bxi) & \propto \prod_{j=1}^{k_m} c(S_j\mid M)g(\bxs_j
      \mid\bxi).
\end{split}
\end{align}
Neither the independence nor the i.i.d. assumption in the first two lines
of \eqref{HierModel} are  strictly   needed for the upcoming discussion,
and could be relaxed.

\section{PPMx with Missing Covariates}\label{sect:PPMxMissing}

% In a prediction setting, missing values
%  occur not only with subjects
% whose observations are employed to fit a particular statistical model, but
% also when predicting for a new (out-of-sample) subject. An additional
% challenge arises when the missingness pattern in the predictor profile  of
% a future  subject is new, i.e., has not been seen in any of the subjects
% used to train the model. The methodology we propose naturally accommodates
% this type of situation.

We extend the PPMx model \eqref{HierModel} to allow for variable dimension
covariate vectors. In short, we generalize the similarity function
$g(\bxs_j)$ in the prior for the random partition to use only available
covariates. We introduce this construction next. This will eventually lead
to a variable-dimension covariate regression. We refer to the proposed
model in general, and the implied variable-dimension covariate regression
as VDReg.

\subsection{Random partitions with variable-dimension covariates}

To develop an extension of the PPMx model  that accommodates missing
covariates, denote by $\OO_i$ the collection of covariate indices that are
observed for subject $i$. The $i$th subject's observed covariate vector
can be now denoted as $\bx_i^o = \{x_{i\ell} : \ell \in \OO_i\}$ and the
collection of observed covariate vectors that belong to the $j$th cluster
is $\bx_j^{\star o} = \{\bx^{o}_{i} : i \in S_j\} = \{x_{i\ell} : \ell \in
\OO_i, i\in S_j\}$. Then missing covariates can be accommodated in the
PPMx by evaluating the similarity function $g_\ell$  using only subjects
$i \in \Cjl =\{i:i\in S_j, \ell \in \OO_i\}$, i.e., those with observed
covariate $\ell$.
% $i$ in the $j$th cluster with $x_{i\ell}$ observed.
Letting $\xjlo=\{ x_{i \ell}: i \in \Cjl\}$, we define a modified
similarity function as
\begin{equation}
\tg(\bx_j^{\star o}\mid\bxi)
\defeq  \prod_{\ell=1}^p \tg_\ell(\xjlo \mid \bxi_{ \ell}) \defeq
        \prod_{\ell=1}^p \int \prod_{i\in \Cjl} q(x_{i \ell}  \mid \bzeta_{j \ell})\,
        dq(\bzeta_{j \ell} \mid \bxi_{\ell}).
\label{NewSim}
\end{equation}

Importantly, in the presence of missing covariates, the similarity
function for the $\ell$th covariate is evaluated based only on subjects
for which the covariate is measured. In other words, missing values are
simply skipped over when evaluating the similarity function. As a result,
no imputation (implicit or not) is being employed.
\cite{xu&mueller&tsim&berry:2019} used a similar strategy when using the
PPMx in a basket trial design, but without any notion of prediction. From
a computational viewpoint, the methods needed to fit this model are
unchanged with respect to the case with no missing observations save for a
matrix of indicators must be carried along. General computational details
can be found in \cite{page&quintana:2015} and we provide more specifics in
Section \ref{computation} of the online supplementary material.

We note briefly that in the context of variable selection
\cite{quintana&mueller&papoila:2015} consider similarity functions that
are similar in form to \eqref{NewSim}, but with each cluster selecting a
cluster-specific subset of covariates. Importantly, in that application
$\mathcal{C}_{j\ell}$ is a random cluster-specific parameter that includes
the subset of covariates that were selected for the  $j$th cluster.  In
that case it is important that $g(x_{j\ell}^{\star})$ be scaled such that
$g(x_{j\ell}^{\star})>1$ for $x_{j\ell}^{\star}$ that are judged to be
very similar and $g(x_{j\ell}^{\star})<1$ for very diverse
$x_{j\ell}^{\star}$.  That is, $g(\cdot)$ needs to be centered around 1,
lest it would introduce an inappropriate prior probability
for including a variable.  %  unrelated to $\bxs_j$.
\cite{quintana&mueller&papoila:2015} introduce an additional factor to
ensure such scaling. However, this issue does not arise here,  since
$\CC_{j\ell}$ is fixed, i.e., inference is conditioned on the observed
covariates.

\subsection{Variable dimension covariate regression (VDReg)}
An important feature of the PPMx prior on partitions is the flexibility in
capturing the role  of covariates in the predictive distribution which we
now discuss.
%
% \subsubsection*{Random partition}
%\mynote{removed the par header here and next page}
The new similarity function in~\eqref{NewSim} easily accommodates
incomplete covariate vectors when making predictions for ``new''
individuals, even if the pattern of missingness has not been observed
among individuals included in the training data set. To see this, consider
the predictive multinomial probabilities  that the $(m+1)$st subject
belongs to one of the groups $h = 1, \ldots, k_m$ conditional on $\rho_m$:
\begin{align}
  p(c_{m+1} = h \mid \rho_m, \bxo, \bx_{m+1}^o ) \propto
  \begin{cases}
    \displaystyle\frac
    {c(S_{h}\cup \{m+1\})\tg(\bx^{\star o}_{h}\cup \{\bx^{o}_{m+1}\})}
    {c(S_{h})\tg(\bx^{\star o}_{h})}
       & \mbox{ for } h = 1, \ldots, k_m\\
    c(\{m+1\})\tg(\{\bx^{o}_{m+1}\})
       & \mbox{ for } h = k_m+1.
  \end{cases}
 \label{eq:predmiss}
\end{align}
where $\tg(\bxso_{h}\cup \{\bxo_{m+1}\})$ is computed including $i=m+1$ in
$S_h$. That is, letting $\Ctjl = \{i:\; i \in S_j \cup \{m+1\} \mbox{ and
} \ell \in \OO_i\}$, we define
\begin{equation}\label{eq:simmissing}
  \tg(\bxso_{h}\cup \{\bxo_{m+1}\}) =
  \prod_{\ell=1}^p\int
   \left\{\prod_{i\in\Ctjl}  q(x_{i \ell }  \mid \bzeta_{h \ell})\right\}~
     dq(\bzeta_{h \ell} \mid \bxi_{\ell}).
  \end{equation}

Thus, any missing covariate for the $(m+1)$st subject is handled
in~\eqref{eq:simmissing} by simply skipping over those missing values, and
therefore, the similarity can be always evaluated. In the extreme case of
a ``new'' subject with an entirely missing covariate vector,  then
$\bxso_{h}\cup \{\bxo_{m+1}\} = \bxso_{h}$ implying that
$\tg(\bxso_{h}\cup \{\bxo_{m+1}\}) = \tg(\bxso_h)$ and thus the
conditional probabilities for the cluster membership indicator in \eqref{eq:predmiss} reduce to
those when making predictions using the PPM.

\begin{figure}[hbt]
  \begin{center}
   \includegraphics[width=.6\textwidth]{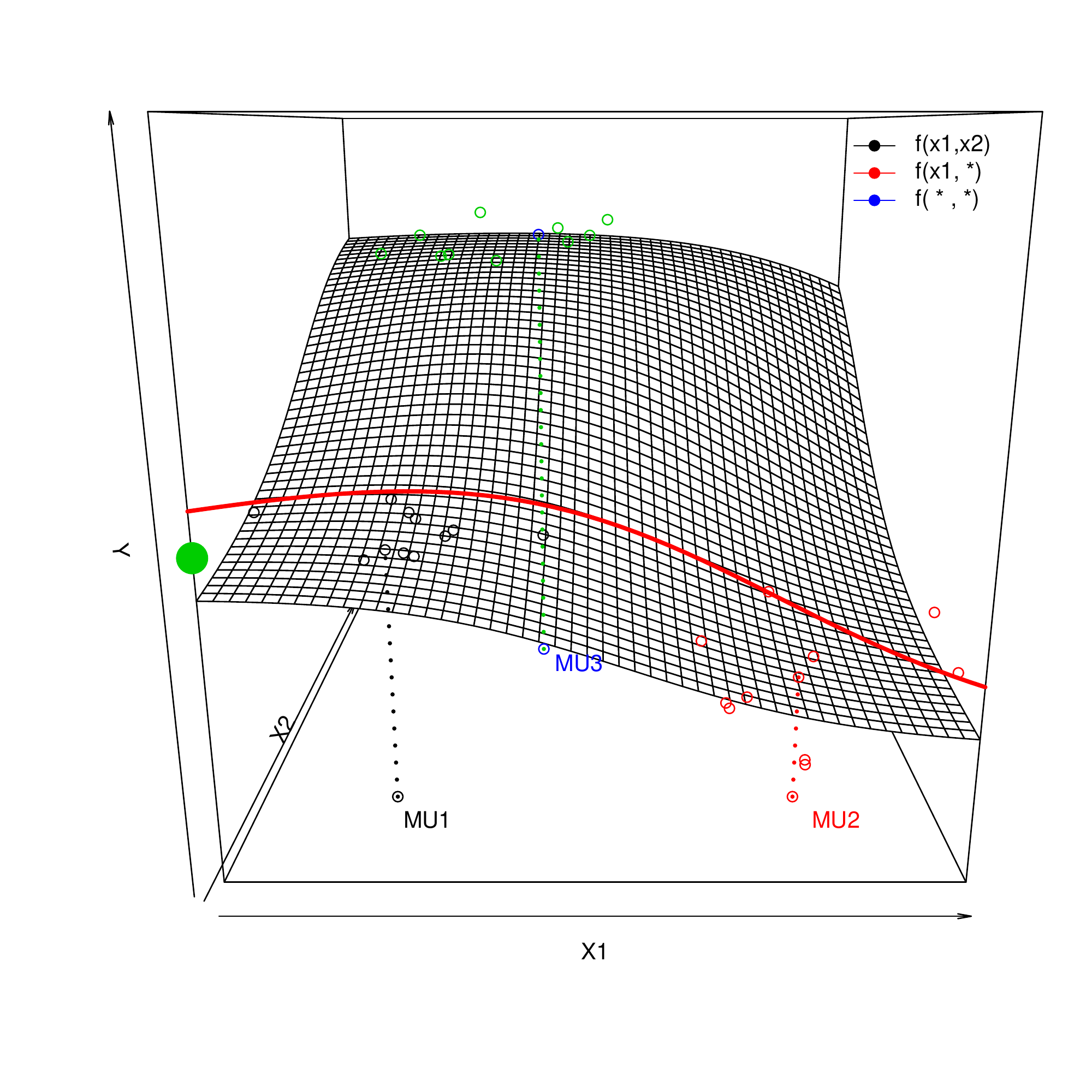}
 \end{center}
\caption{Regression with variable dimension covariates, for data
$(x_{1i},x_{2i},y_i)$ with black, blue and red indicating data linked to
three imputed clusters. The plot shows the posterior predictive regression
$f(x_1,x_2) = E(y_{n+1} \mid x_{n+1,1}=x_1,x_{n+1,2}=x_2, {\rm data})$ for a
future data point with both, $(x_1,x_2)$ observed (black response
surface), $f(x_1,\bullet)$ for a data point $(x_1,\bullet)$ with missing
$x_2$ (red curve) and $f$ for a future observation without available
covariates (green bullet on the Y-axis). }
 \label{fig:ppmx}
\end{figure}

% \subsubsection*{The predictive model - variable dimension covariate
% regression}
To allow for prediction we add the sampling model \eqref{HierModel} (first
two lines) to include responses $y_i$.
% At this moment we have to restrict
% $p(y_i \mid \bths_j, c_i=j, \bx_i)$ to
% $p(y_i \mid \bths_j, c_i=j)$, without additional regression on
% covariates in the sampling model. The restriction is minor
% since regression on $\bx_i$ is already
% provided in \eqref{eq:predmiss}.
In the full model, posterior predictive probabilities \eqref{eq:predmiss}
for the cluster membership $c_{m+1}$ imply a flexible regression for
$y_{m+1}$ on $\bxo_{m+1}$. In words, the regression is described as a
locally weighted mixture of predictions under different clusters, with the
local weighting induced by \eqref{eq:predmiss} and all being marginalized
with respect to posterior uncertainty on the clustering. The local
weighting introduces the regression on $\bxo_{m+1}$, with the desired
feature of allowing variable dimension $\bxo_{m+1}$. This is because only
observed covariates are used in \eqref{eq:predmiss}. Formally, let
$w_j(\bxo_{m+1};\; \bxso_j)=p(c_{m+1} = h \mid \rho_m, \bxo, \bxo_{m+1})$,
and let $f_j(y_{m+1};\; \bys_j) = \int p(y_{m+1} \mid c_{m+1}=j, \bys_j,
\bths_j)\, dp(\bths_j \mid \bys_j)$. We get a locally weighted regression
\begin{equation}
  p(y_{m+1} \mid \bx, \by, \bx_{m+1}, \rho_m) =
  \sum_{j=1}^{k_m} w_j(\bx_{m+1};\; \bxs_j)\, f_j(y_{m+1}; \bys_j)
\label{eq:pymrho}
\end{equation}
and finally
\begin{equation}
p(y_{m+1} \mid \bxo, \by, \bxo_{m+1}) =
 E_\rho\left\{
  \sum_{j=1}^{k_m} w_j(\bx_{m+1};\; \bxs_j)\, f_j(y_{m+1}; \bys_j)
  \mid \by \right\}.
\label{eq:pym}
\end{equation}
Expression \eqref{eq:pymrho} clearly exposes how the variable dimension
covariate regression is implemented by using weights $w_j$ that only make
use of  available covariates. The final equation averages over the unknown
partition, with respect to $p(\rho_m \mid \by, \bxo)$. The regression
\eqref{eq:pym} concisely summarizes the proposed approach to implement
variable dimension covariate regression.
% The statement in \eqref{eq:pymrho} also highlights again an important
% limitation of he setup. There is no adjustment for possible missing
% not at random, i.e., for possible information
In summary, by the implied posterior predictive distribution
\eqref{eq:pym} the proposed model defines a  variable dimension covariate
regression. This is illustrated  in Figure \ref{fig:ppmx}.

\subsubsection*{PPMx with missing covariates}

For the upcoming simulation study and ozone data example we implement the
model for continuous outcomes $y_i$ and continuous covariates $x_{i\ell}$,
using a normal sampling model $p(y_i \mid c_i=j,
\bm{\theta}^{\star}_j=(\mus_j, \sigs_j)) = N(\mus_j, \sigs_j)$ with a
conjugate normal prior on the location parameter and a uniform prior on
cluster-specific standard deviations $\sig^\star_j$. For the cohesion
function we use $c(S_j \mid M) = M (|S_j|-1)!$. The similarity functions
are specified using  \eqref{NewSim}, with $q(x_{i\ell} \mid
\bzeta_{j\ell})$ and $q(\bzeta_{j\ell} \mid \bxi)$  with fixed values for
$\bxi$ and other hyperparameters. For later reference we summarize the
complete VDReg model with these choices:
\begin{align}\label{HierModelNorm}
\begin{split}
  y_i \mid \bm{\mu}^*, \bm{\sigma}^{2*},  {c_i=j} &
  \sim N(\mu^*_{j}, \sigma^{2*}_{j}) \ \mbox{for $i = 1, \ldots, m$}\\
(\mu^*_{j}, \sigma^{\star}_j \mid  \rho) & \sim N(\mu_0, \sigma^2_0) \times Uniform(0, a_{\sigma}) \ \mbox{for $j = 1, \ldots, k_m$}\\
(\mu_0, \sigma_0) & \sim N(m_0, v^2) \times Uniform(0, a_{\sigma_0}) \\
p(\rho=\{S_1, \ldots, S_{k_m}\}\mid\bx^{o},M,\bxi) &
\propto \prod_{j=1}^{k_m} c(S_j\mid M)\tg(\bx^{\star o}_j\mid\bxi).
\end{split}
\end{align}
The uniform prior on cluster-specific standard deviations follows
suggestions in \cite{gelman2006}.
% We will refer to
% \eqref{HierModelNorm} as ``PPMxM" or PPMx with missing covariates.
% \footnote{ We state this earlier.  Should we remove it from the last sentence in the first paragraph of Section 3? }
The function $\tg(\cdot)$ in the last line is where the model accommodates
variable-dimension covariate vectors using \eqref{eq:simmissing} to define
a similarity function on the basis of available covariates only. This is
at the heart of the proposed VDReg model.

% \section{Simulation Study} \label{simstudy.and.data}

% In this section we describe a simulation study designed to explore the
% performance of our proposed method with regards to model fit and
% out-of-sample prediction.  Then we consider a real-world data set
% containing ozone measurements.

\section{Simulation Study}\label{simulation.study}

We conduct a simulation study to assess  how  predictions are affected by
(1) an increase in the number of covariates and the missingness  rate, (2)
different types of missingness, and (3) how distinct covariates are across
different clusters.

\paragraph*{Simulation scenarios.}
We generate  synthetic data with 100 testing and 100 training
observations.  We describe  next   the generation of covariates, responses
and the missing data.
\smallskip

\noindent \underline{Covariates:} Data sets are generated  with  a varying
number of covariates, $p \in \{2, 4, 10\}$.  Specific values for the
covariates are generated using four $p$-dimensional Gaussian
distributions, thus creating $k_m=4$ covariate dependent clusters. For
example, when $p=2$, we use four bivariate normals, $N(m_j, V_j)$, with
$m_j = (1,1)$, $(1,-1)$, $(-1, 1)$, and $(-1,-1)$ to generate 200 sets of
covariate values $\bx_i$ (50 in each cluster). Similarly, with $p=4$, we
use  $m_j=(1,1, 1, 1)$, $(1,-1, 1, -1)$, $(-1, 1, -1, 1)$, and $(-1,-1,
-1, -1)$, to create $k_m=4$ clusters with 50 observations each. And
lastly, for $p=10$, we use $m_j=(1,1, 1, 1,1,1,1,1,1,1)$, $(1,-1, 1,
-1,1,-1, 1, -1, 1, -1)$, $(-1, 1, -1, 1,-1, 1, -1, 1, -1, 1)$, and
$(-1,-1, -1, -1,-1,-1, -1, -1, -1, -1)$. Notice that as $p$ increases
$k_m=4$ remains constant, but the clusters become  sparser  in the
covariate space. To study how the  overlap among clusters (``cluster
noise'') affects prediction results, the covariance matrix $V_j$ used to
generate covariates is set as $V_j=s^2\bm{I}_p$ with $s^2 \in \{0.25^2,
0.5^2, 0.75^2\}$. Under $s^2=0.25^2$ the clusters are well separated,
while for $s^2=0.5^2$ the clusters  are adjacent,  and for $s^2=0.75^2$
the clusters overlap  substantially. Figure \ref{cluster.config} in the
online supplementary material displays the cluster configuration for each
of the $s^2$ when $p=2$. This describes the generation of the covariates
$\bx_i$.
\smallskip

\noindent \underline{Responses} $y_i$  are generated as $p(y_i \mid
s_i=j)=N(\mus_j, \sigs_j)$. We use $\mus_j=-1, -0.5, 0$, and $0.5$  for
observations in clusters $j=1$ through $4$, respectively. To study how
heterogeneity of variances across clusters impacts inference we use two
sets of simulations, one with $\sigs_j=0.25^2$ for all $j$, and one with
$\sig^\star_j=0.1, 0.25, 0.5$ and $0.75$, respectively for $j=1,\ldots,4$.
% first sets the standard deviation of each Gaussian equal to 0.25 and the
% second uses $(0.1, 0.25, 0.5, 0.75)$ as standard deviations in the four
% Gaussian distributions.
\smallskip

\noindent \underline{Missing values} in the covariates are inserted as
follows. For each covariate a specific fraction (approximately) of values
are randomly selected to be classified as missing. We consider two types
of missingness. The first is missing  at random (MAR) and the second
missing not at random (MNAR). Generating both types of missing is
facilitated using the {\tt ampute} function found in the {\tt mice}
R-package (\citealt{micePack}). For MNAR, the {\tt ampute} function  is
used for each covariate with the missing probabilities being a function of
the covariate value (see \citealt{schouten&lugtig&vink} for specific
details regarding the function used to produce probability of missing).
The {\tt ampute} function is also applied separately to each covariate for
the MAR case where each covariate entry is equally likely to be classified
as missing. In summary, we generate data under simulation truths varying
% numerical experiment considers
the following factors:
% \begin{enumerate}
(1) {\em type of missing} (MAR or MNAR),
(2) {\em missing fraction} (0\%, 10\%, 25\%, 50\%),
(3) {\em number of  covariates} ($p=2, 4$ and $10$),
(4) {\em cluster noise} ($s^2 \in \{0.25^2, 0.5^2, 0.75^2\}$), and
(5) {\em heteroscedasticity} (yes, no).

\paragraph*{Comparison.}
Each  created synthetic data set is comprised of 200 observations (50 in
each cluster).  Then the synthetic datasets are randomly split into 100
training and 100 test observations. For each simulated data set we
implement inference under the following models and approaches:
%\begin{enumerate}\itemsep=0pt
(1) BART: The method detailed in \cite{kapelner&bleich:2015} and
    carried out using the {\tt bartMachine} package
    (\citealt{kapelner&bleich:2016}) in {\tt R};
(2) MI: Based on 10 imputed data sets via the {\tt complete}
    function of the {\tt mi} package (\citealt{gelman&hill:2011}) from
    the statistical software {\tt R} (\citealt{Rmanual:2018});
(3) PSM:  Pattern submodel approach using  method in
    \cite{mercaldo&bluem:2018} and code available at
    \url{https://github.com/sarahmercaldo/MissingDataAndPrediction}.\\
(4) VDReg:  model \eqref{HierModelNorm};  and
% \end{enumerate}
When fitting the VDReg model \eqref{HierModelNorm} covariates are
standardized to zero mean and unit variance. For similarity
$g(\bx^{\star}_j\mid\bxi)$ we use \eqref{sf} with
$q_\ell(\cdot\mid\zeta_{\ell j}) = N(\cdot; \zeta_{\ell j}, 0.5)$ and
$q(\zeta_{\ell j}) = N(\zeta_{\ell j}; 0, 1)$.  Finally, fixed
hyperparameters are $m_0=0$, $v^2=100^2$, $a_{\sigma}=10$ and
$a_{\sigma_0} = 10$. With these prior specifications, we fit model
\eqref{HierModelNorm} by collecting 1000 MCMC samples after discarding the
first 25,000 as burn-in and thinning by 25 (i.e., 50,000 total MCMC draws
are sampled).  All computation for model \eqref{HierModelNorm} is carried
using the {\tt ppmx.missing} function that is part of the {\tt ppmSuite}
R-package found on the first author's webpage.

In order to make out-of-sample predictions using MI, covariates in
training and testing data were joined, and imputation was carried out
based only on this joined matrix (i.e., the response associated with
training data was not included in the imputation). Default parameter
values for the BART and PSM procedures are used.

We use the following metrics for the comparison:
% \begin{itemize}\itemsep=0pt
MSE (mean squared error) measures goodness-of-fit,
    $MSE=\frac{1}{100}\sum_{i=1}^{100} (Y_{oi} - \hat{Y}_{oi})^2$, where
    $i$ indexes the 100 training observations ($Y_o$) and
    $\hat{Y}_{oi}$ is the fitted value for the $i$ observation;
and MSPE (mean squared prediction error)     measures the predictive
performance of the models,
    $MSPE =\frac{1}{100}\sum_{i=1}^{100} (Y_{pi} - \hat{Y}_{pi})^2$, where
    $i$ indexes the 100 testing observations $(Y_{p})$ and
    $\hat{Y}_{pi} = E(Y_{pi} \mid \bm{Y}_o)$.

\begin{figure}[htbp]
\begin{center}
\includegraphics[scale=0.5]{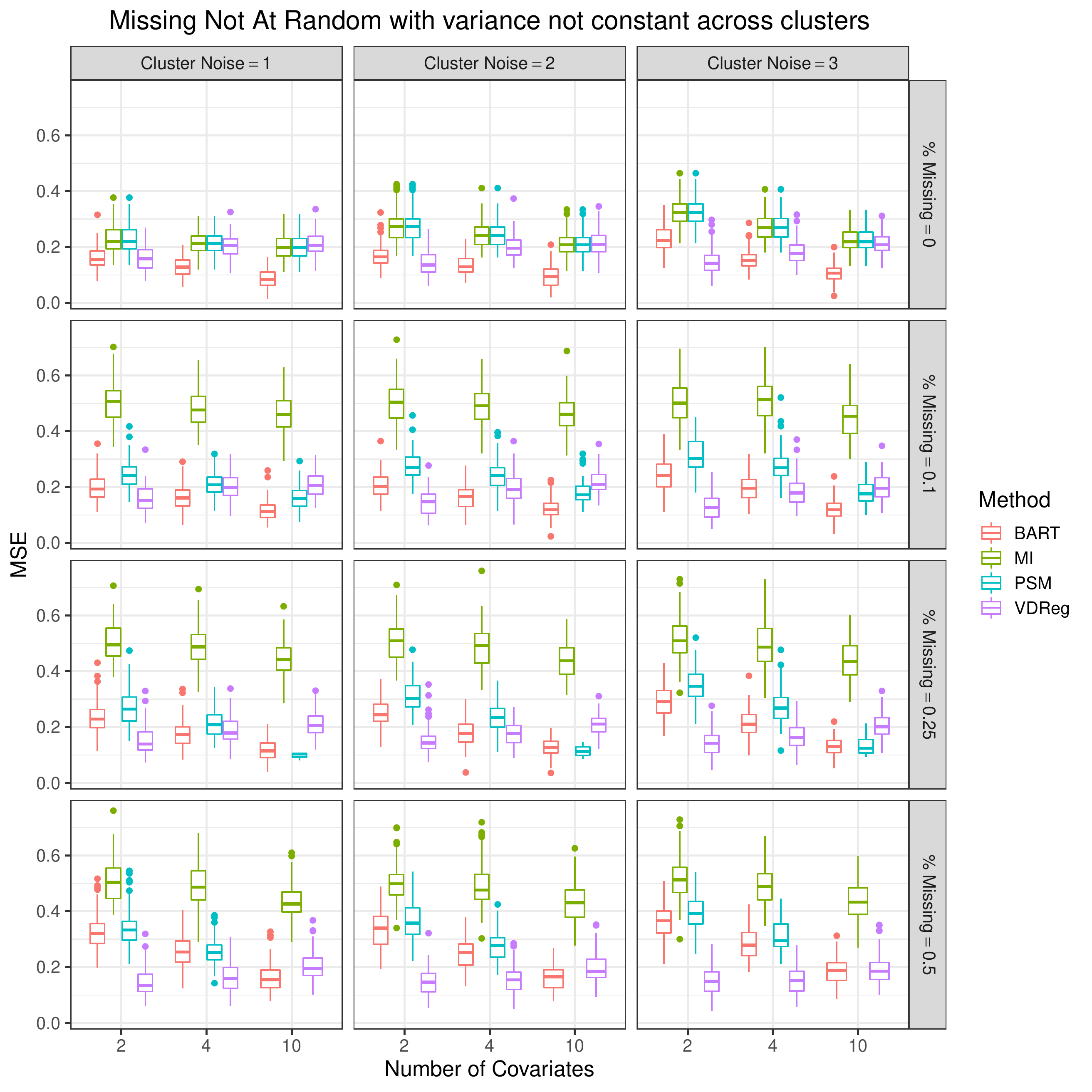}
\caption{MSE results from simulation study when missing is not at random and the
$s^2$ is not constant across clusters.} \label{MSE_MNAR}
\end{center}
\end{figure}

\begin{figure}[htbp]
\begin{center}
\includegraphics[scale=0.5]{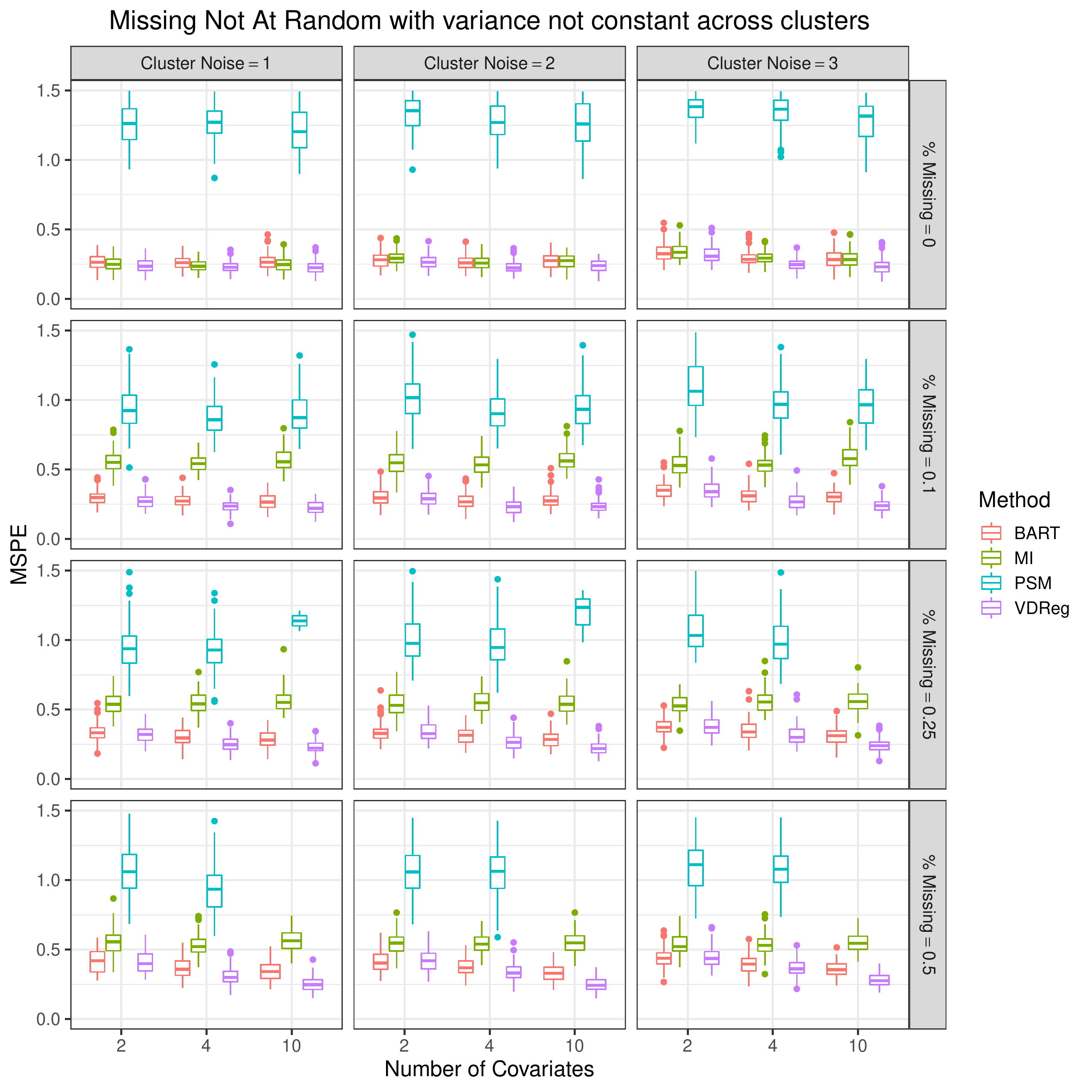}
\caption{MSPE results from simulation study when missing is not at random and the
$s^2$ is not constant across clusters.} \label{MSPE_MNAR}
\end{center}
\end{figure}

\paragraph*{Results.}
Before describing simulation results, we note that under a missing rate of
50\% and $p=10$ covariates the software used to fit the PSM model  exceeds
an internal computational limit and aborts in error. As a result, PSM is
not included in this scenario. We found that the simulation results are
similar under the various combinations of data being MNAR or MAR, and
whether or not we assume heteroscadasticity. We therefore present only
results under MNAR and heteroscedasticity, and summarize other results in
Section \ref{sim.study} of the online supplementary material. Figures
\ref{MSE_MNAR} and \ref{MSPE_MNAR} display MSE and MSPE as a function of
the number of covariates, missing fraction and cluster noise.

Focusing on the MSE values first, notice that under 0\% missing fraction,
BART fits the data best and the other procedures are similar with cluster
noise impacting MI and PSM  the most (which is to be expected). However,
with increasing missing rate VDReg reports the best model fit, with the
differences between procedures increasing with higher missing fraction,
cluster noise and $p$. Generally speaking, MI tends to perform  least
favorably (as one might expect  of   a very generic method).

Regarding MSPE, results are very similar across procedures (save PSM) when
there are no missing values, with relative performance under VDReg looking
increasingly better with increasing cluster noise and $p$. PSM and MI are
at an inherent disadvantage as no linear model was explicitly included in
the top level sampling model. With increasing missing rate the prediction
accuracy of the PSM and MI degrades the most (which was expected). VDReg
and BART generally predict better as the number of covariates increases.
Overall, VDReg is least impacted by an increase in the missing fraction
and cluster noise. The simulation study indicates that VDReg  performs
favorably  in accommodating missing values relative to BART, MI, and PSM,
regardless of the type of missingness (see Figures \ref{MSE_MAR_NonConst}
- \ref{MSPE_MNAR_Const} in the online supplementary material).

\section{Application Examples}\label{sect:data}
\subsection{Ozone Data}\label{sect:ozone}

\begin{figure}[htbp]
\begin{center}
\includegraphics[width=\columnwidth]{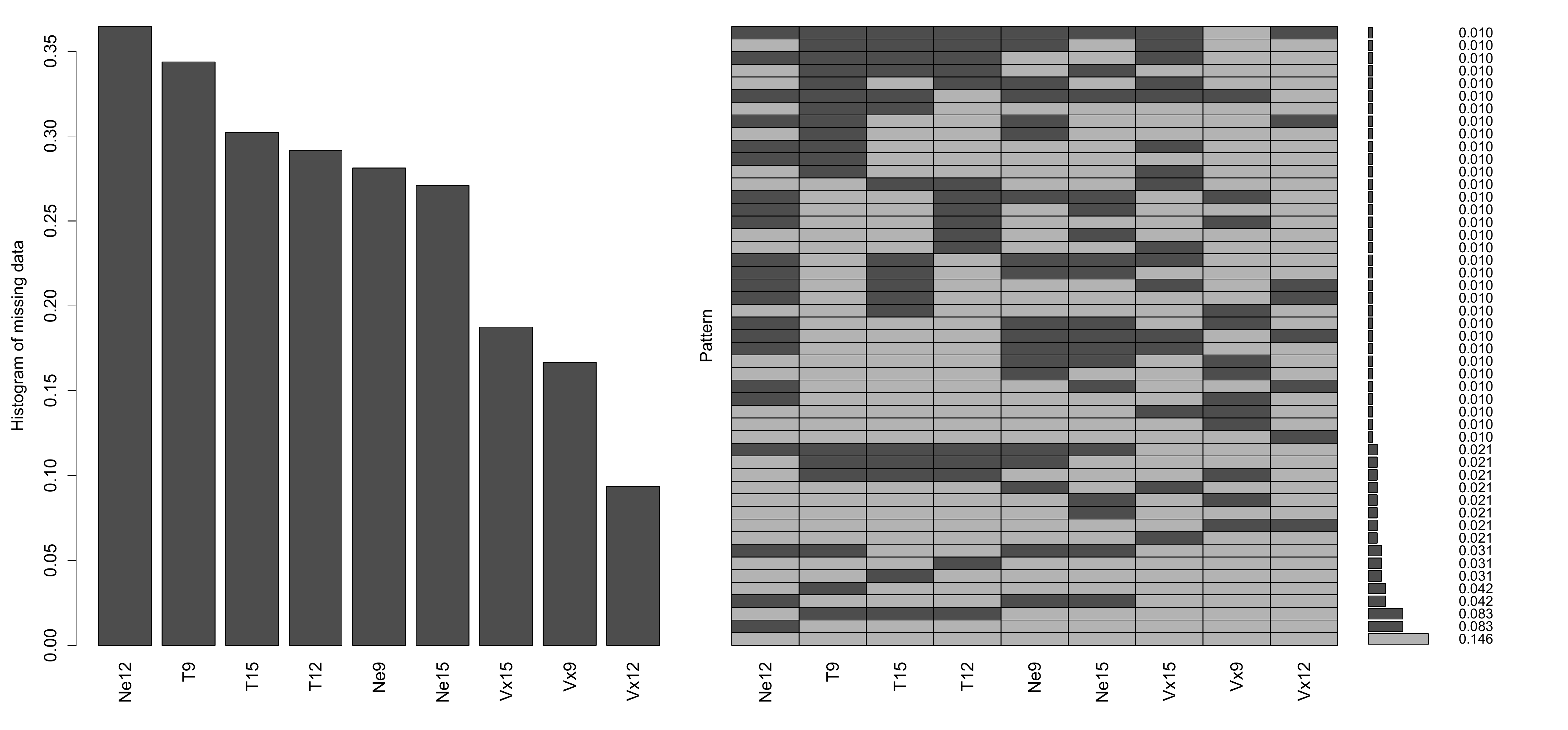}
\caption{Missing rates and patterns associated with the ozone data set. The left
plot displays the percent missing for each covariate. In the right plot
each row corresponds to a missing pattern with cells colored in dark gray
indicating the covariate is missing} \label{ozoneMissingPattern}
\end{center}
\end{figure}

We consider a small environmental data set that is publicly
available.\footnote{https://github.com/njtierney/user2018-missing-data-tutorial/blob/master/ozoneNA.csv}
This data set consists of 112 measurements of maximum daily ozone in
Rennes.  In addition,  temperature (T), nebulosity (Ne), and projection of
wind speed vectors (Vx) were measured three times daily (9:00, 12:00, and
15:00 hours) resulting in nine covariates. There are 16 locations for
which the response (maximum daily ozone measurements) is missing. This
could be handled with any of the existing methods in the literature
focused on missing responses. However, for the sake of simplicity we
remove these observations. Figure \ref{ozoneMissingPattern} displays the
amount of missing for each covariate and the missing patterns. Notice that
there are a number of missing patterns that appear only one time and only
14.6\% of observations are complete cases.

The 96 observations are divided in training and test datasets by randomly
selecting 75 observations as training data and treating the remaining 21
as test data. The procedure of randomly  splitting into training and test
data is repeated 100 times, and each time we fit the training data and
make predictions for the testing data using BART, MI,  PSM, and VDReg
(see the previous section for a brief description of the methods). For
each of the fits MSE and MSPE is calculated. Also, in order to further
study how increasing $p$
% (and as a result the number of missing patterns)
impacts the out-of-sample prediction performance
% of our approach,
we  repeat the described process again using only $p=2$ covariates
(temperature at 9:00 and 12:00), then $p=3$ (temperature at 9:00, 12:00
and 15:00),  and next sequentially adding nebulosity and then projection
of wind speed vectors for each time during the day. The MCMC  details  and
prior values for model \eqref{HierModelNorm} are as in the simulation
study.
% in the simulation study of Section \ref{simulation.study}.
Also as in the simulation study, BART and PSM are fit using the default
tuning parameter values.

\begin{figure}[htbp]
\begin{center}
\includegraphics[scale=0.5]{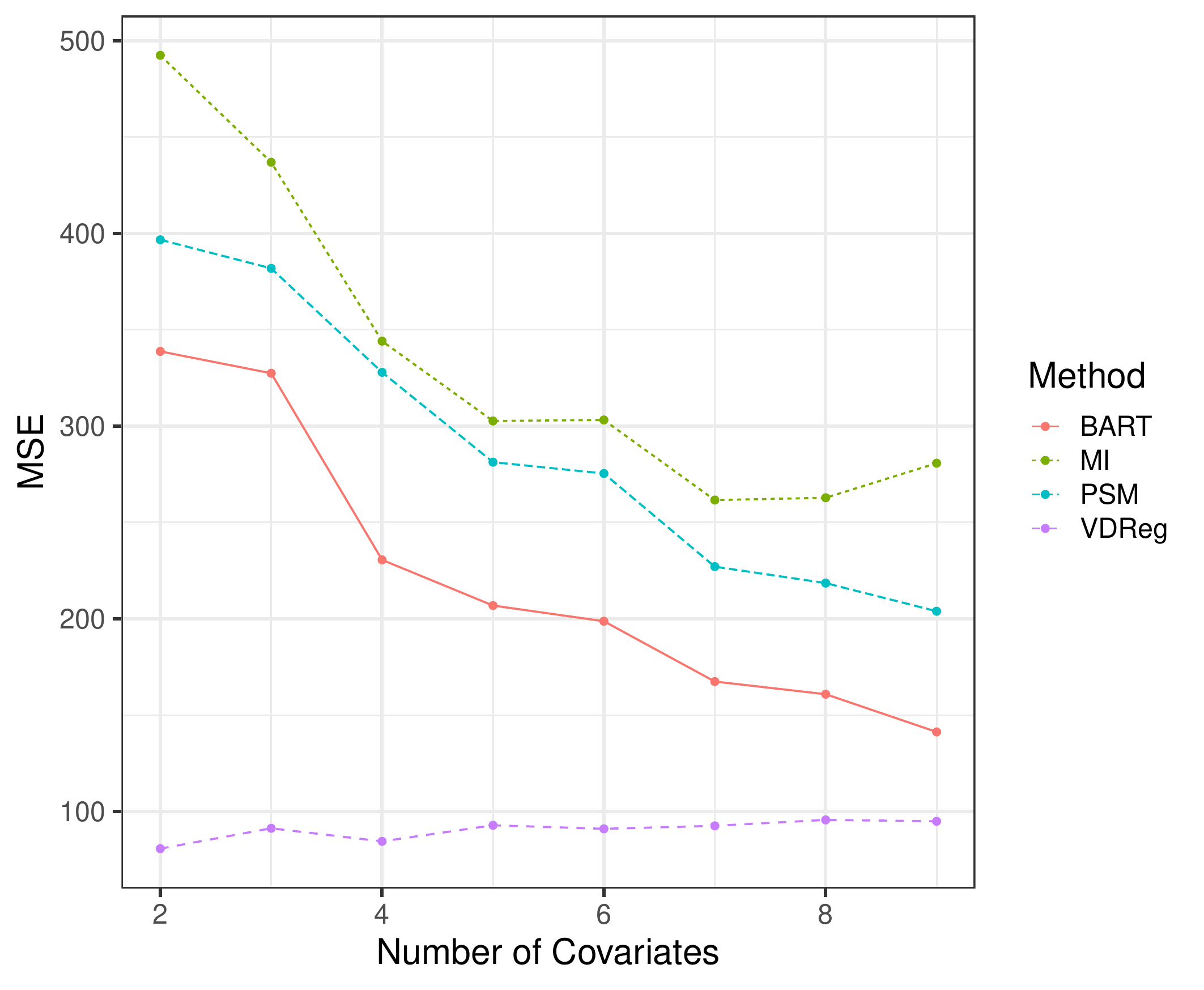}
\caption{MSE values averaged over 100 cross-validation datasets based on ozone data}
\label{OZONEmse}
\end{center}
\end{figure}

\begin{figure}[htbp]
\begin{center}
\includegraphics[scale=0.5]{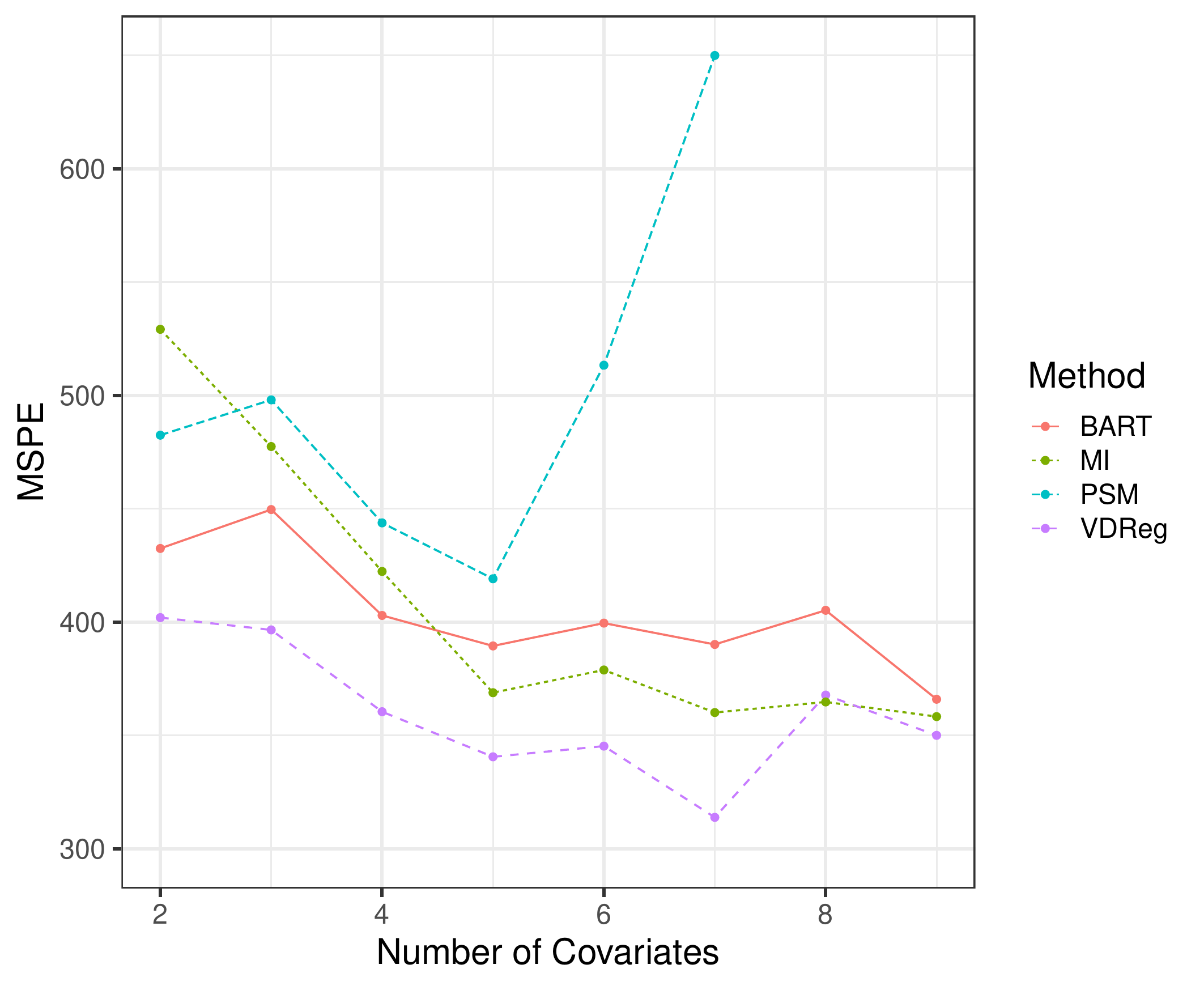}
\caption{MSPE values averaged over 100 cross-validation datasets based
  on ozone data
  (values for PSM for $p=8$ and $9$ are beyond the plotting limits). }
\label{OZONEmspe}
\end{center}
\end{figure}

The average MSE and MSPE values over the 100 cross-validation data sets
are provided in Figures \ref{OZONEmse} and \ref{OZONEmspe}. From Figure
\ref{OZONEmse}  notice that the MSE values for the VDReg model are lower
than BART,  MI, or PSM regardless of the number of covariates that are
considered. In terms of out-of-sample prediction, it seems that VDReg has
the lowest MSPE among the five methods regardless of the number of
covariates. It seems that the PSM method performed the worst with
performance decaying drastically as the number of covariates is increased.
% (Note that graphical purposed MSPE values from the PSM procedure were
% excluded when the model included 8 or 9.)

\subsection{Prostate Cancer Data}
We consider data  from a prostate cancer study that was analyzed in
\cite{deng_etal:2016},  who  employ two variations of imputation to
accommodate missing covariates.  The data set is publicly available (GEO
GDS3289) and is based on 99 subjects, including 34 benign and 65 produce
malignant epithelium samples, each with 20,000 biomarkers. We code  the
response as $y_i=1$ for benign samples and $y_i=0$ for malignant. Besides
a minor adaptation of \eqref{HierModelNorm} for the binary response in the
sampling model, the VDReg model can be  employed without alterations.
% Among the many options that are available to model binary response
% data,
We use a  latent probit score, i.e., $p(y_i=1 \mid c_i=j, \mus)=
\Phi(\mus_j)$ (with $\Phi(\cdot)$ denoting a standard normal c.d.f.), and
otherwise leave \eqref{HierModelNorm} unchanged. Of the 20,000 biomarkers,
\cite{deng_etal:2016} focused on three (FAM178A, IMAGE:813259 and UGP2)
that are known to be associated with the response.  The missing rates of
the three covariates are 31.3\%, 45.5\%, and 26.3\% respectively.
\cite{deng_etal:2016} then use multiple imputation methods based on 2107
biomarkers that do not have any missing values and then using the imputed
datasets, fit a logistic regression model and report estimates of the
regression coefficients.

Since our focus is on prediction, we instead split the 99 subjects into 75
training and 24 testing observations and fit the VDReg model.
  % based on a probit regression data model.
Splitting the data set into training and testing observations was carried
100 times and  for each split we evaluated within sample and out of sample
predictions.  Results are shown in Table \ref{prostate}.  In addition to
prediction rates we report Tjur's $R^2$ (\citealt{tjur:2009}).  This
metric compares the average estimated probability of being in the benign
group for subjects with benign samples to the average estimated
probability of being in the malignant group for subjects with malignant
samples.  As this number approaches one, it is an indication of superior
model fit. For comparison, we also include results under BART and PSM (as
in Section~\ref{sect:ozone}). VDReg compares favorably to the other two
methods in terms of in-sample prediction rate and Tjur's $R^2$ value.  For
out-of-sample prediction VDReg does slightly better than the other two
methods, but with worse Tjur's $R^2$.

\begin{table}[htp]
\caption{Cross validation results based on the prostate cancer data. Each
of the 100 cross-validation data sets were comprised of 75 training and 24
testing observations. Results presented in the table are averages over the
100 cross-validation datasets. }
\begin{center}
\begin{tabular}{l|cccc}
& \multicolumn{2}{c}{In Sample Prediction} & \multicolumn{2}{c}{Out Sample Prediction}\\ \cmidrule(r){2-3} \cmidrule(r){4-5}
Method & \% Correct & Tjur $R^2$ &  \% Correct & Tjur $R^2$\\ \midrule
BART 		& 0.81 & 0.29 & 0.70 & 0.16 \\
PSM 		& 0.79 & 0.39 & 0.70 & 0.24 \\
VDReg 		& 1.00 & 0.63 & 0.71 & 0.16 \\
% BART 		& 0.8092 & 0.2876 & 0.6975 & 0.1648 \\
% PSM 		& 0.7920 & 0.3914 & 0.7044 & 0.2436 \\
% VDReg 		& 0.9997 & 0.6228 & 0.7096 & 0.1561 \\
\end{tabular}
\end{center}
\label{prostate}
\end{table}%

Lastly,   % as a means to compare the VDReg to the imputation methods
by way of comparison with the imputation methods used in
\cite{deng_etal:2016}, using the estimated logistic regression
coefficients reported in \cite{deng_etal:2016} we predicted cancer status
for  the 26 complete cases found in the data set.  We then fit VDReg to
all 99 observations and also predicted the cancer status of the 26
complete cases.  Of these 26 predicted outcomes, the VDReg was correct for
$88\%$ of them compared to $69\%$ based on the imputation methods.
% ith a true positive rate of 0.85 and a true negative rate of 1.0.
% While predictions based on the coefficients estimated using
% imputation methods described in \cite{deng_etal:2016} produced a
% prediction with 0.69, a true positive rate of 0.62 and a true
% negative rate of 1.0.

\section{Conclusions}\label{sect:concl}

We have extended the PPMx random partition model to allow for missing
covariate values without resorting to any imputation or substitution. This
is particularly useful when the main inferential target is prediction. The
proposed  approach facilitates out-of-sample predictions with any subset
of covariates.

Some limitations remain, and provide opportunities for further
generalizations. In the current form the model does not include any notion
of variable selection or transformation. While independent variable
selection is straightforward to add, the use of partially missing
covariate vectors would complicate any approach that involves dependent
priors over variables. Similarly, the use of any transformation or
projections of the joint covariate vector is not straightforward in the
presence of missing covariates without imputation. In preliminary results
not shown,  we explored the proposed method in the case when the
underlying data structure is such that only a small number of covariates
inform the partition relative to the total number measured. We found that
the PPMx model in these circumstances is not as competitive as the BART
approach, as indicated in our simulation study. In the case of a scenario
with many covariates, we suggest first employing some dimension reduction
or variable selection technique (one option is described in
\citealt{page2018discovering}), and afterwards applying our approach based
only on those covariates that are useful.

\section*{Acknowledgments}

Garritt L. Page acknowledges support from the Basque Government through the BERC 2018-2021 program, by the Spanish Ministry of Science, Innovation and Universities through BCAM Severo Ochoa accreditation SEV-2017-0718
F. Quintana's research is supported by Millennium Science Initiative of
the Ministry of Economy, Development, and Tourism, grant ``Millennium
Nucleus Center for the Discovery  of Structures in Complex Data''. F.
Quintana is also supported by Fondecyt grant 1180034.  P. M\"uller acknowledges partial support from grant NSF/DMS 1952679
from the National Science Foundation, and under R01 CA132897 from the
U.S. National Cancer Institute. 

\singlespace
 \bibliographystyle{dcu}
\bibliography{reference}

\end{document}

\onehalfspace

\title{\textbf{Supplementary material}\\Prediction in the Presence of Missing Covariates}
\author{ Garritt L. Page \\ Department of Statistics \\ Brigham Young University, Provo, Utah \\  and \\ Basque Center of  Applied Mathematics \\page@stat.byu.edu
         \and	Fernando A. Quintana \\ Departamento de Estad\'{i}stica \\ Pontificia Universidad Cat\'{o}lica de Chile, Santiago
         \\ and Millennium Nucleus Center for the \\ Discovery of Structures in Complex Data \\ quintana@mat.puc.cl
        \and Peter M\"uller \\ Department of Mathematics \\ The University of Texas at Austin \\ pmueller@math.utexas.edu
}

\maketitle

\doublespace This document contains details associated with computation
needed to fit model \eqref{HierModelNorm} along with additional results
from the simulation study provided in the paper ``Prediction in the
Presence of Missing Covariates''.

\section{Computation} \label{computation}
Fitting the models detailed in equation \eqref{HierModelNorm} of the main
article requires constructing an MCMC algorithm that samples from the
joint posterior distribution of  model parameters. The only real
difference between an algorithm for complete data and that for missing
data is that an indicator matrix that identifies which covariate values
are missing needs to be carried along when updating cluster labels.    The
algorithm we employ is a hybrid Gibbs-sampler with Metropolis steps that
is based on algorithm 8 of \cite{neal:2000}.  The variance components
$(\sigma^{2\star}_1, \ldots, \sigma^{2\star}_{k_m}, \sigma^2_0)$ are
updated using random walk Metropolis steps with a Normal distribution used
to generate candidate values.  The means/intercepts for both models
$(\mu{\star}_1, \ldots, \mu^{\star}_{k_m}, \mu_0)$ are updated using a
Gibbs step as their full conditionals are Normal and can be derived using
well known arguments.  Updating $\bm{\beta}_{\ell}$ is also done using a
Gibbs step as its full conditional is a multivariate normal which can also
be derived using well known arguments.  The only difference is that care
must be taken that only observations with a specific missing pattern be
used.  Updating the cluster labels is also very similar to that based on
complete data case save $\tilde{g}(\cdot)$ is used.  In detail, the $i$th
cluster label in model \eqref{HierModel} is updated using the following
multinomial probabilities.  For $h=1, \ldots, k_m^{-i}$ (where $k_m^{-i}$
denotes the number of clusters after having removed the $i$ observation)
\begin{align}
Pr(c_{i} = h | - ) \propto
  \begin{cases}
    N(y_i; \mu^{\star}_h, \sigma^{2\star}_h) \displaystyle\frac{c(S^{-i}_{h}\cup \{i\})\tilde{g}(\bm{x}^{\star o (-i)}_{h}\cup \{\bm{x}^{o}_{i}\})}{c(S^{-i}_{h})\tilde{g}(\bm{x}^{\star o (-i)}_{h})}\ \mbox{for} \ h = 1, \ldots, k_m^{-1}\\
    N(y_i; \mu^{\star}_{new, h}, \sigma^{2\star}_{new, h}) c(\{i\})\tilde{g}(\{\bm{x}^{o}_{i}\})  \ \mbox{for} \  h = k_m^{-1}+1,
\end{cases}\label{eq:predmiss}
\end{align}
where $\mu^{\star}_{new, h}$ and $\sigma^{2\star}_{new, h}$ are drawn from their respective prior distributions.

\section{Additional Results from Simulation Study}\label{sim.study}
\begin{figure}[htbp]
\begin{center}
\includegraphics[scale=0.45]{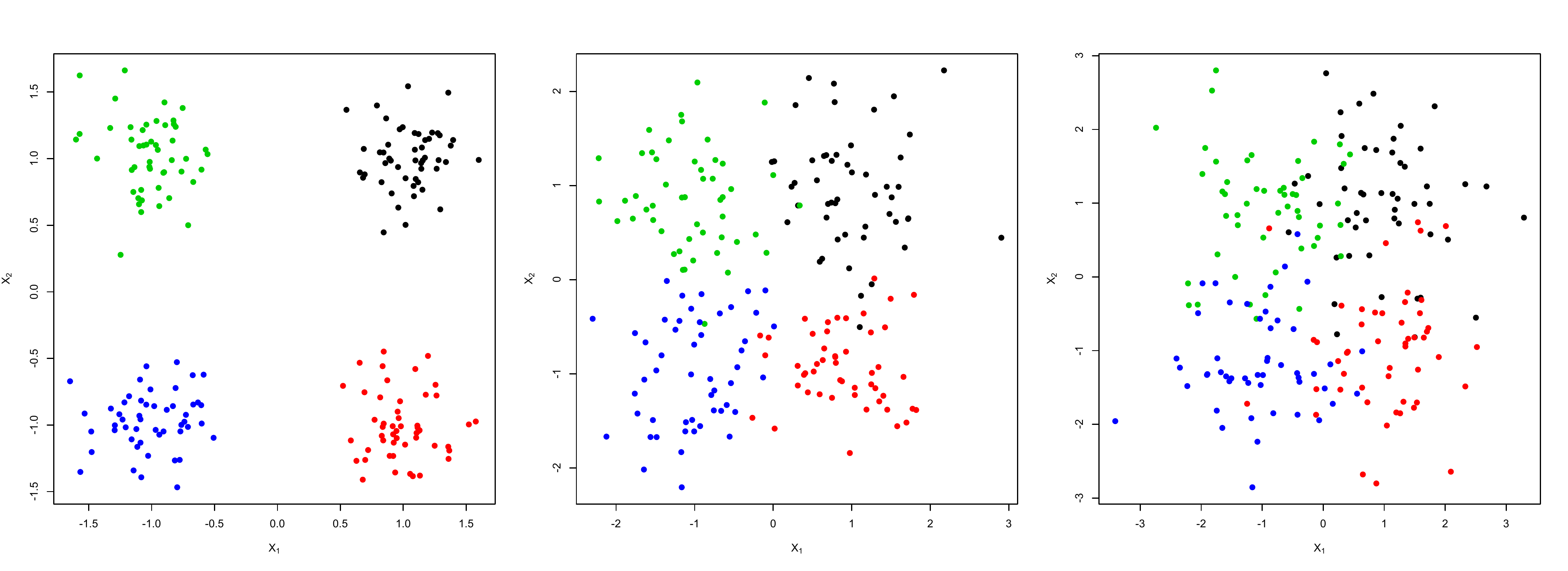}
\caption{Cluster configuration employed in the simulation study for $p=2$ and for $s^2=0.25^2$ (left plot), $s^2=0.5^2$ (center plot), and $s^2=0.75^2$ (right plot).}
\label{cluster.config}
\end{center}
\end{figure}

In Figure \ref{cluster.config} we provide an example of the cluster
configuration when $p=2$.  Notice that as $s^2$ increases, the clusters
become more ``noisy''.  The thing to notice in these plots is that as
$s^2$ increases, the overlap between observations belonging to different
clusters also increases.  This algorithm is a

\begin{figure}[htbp]
\begin{center}
\includegraphics[scale=0.45]{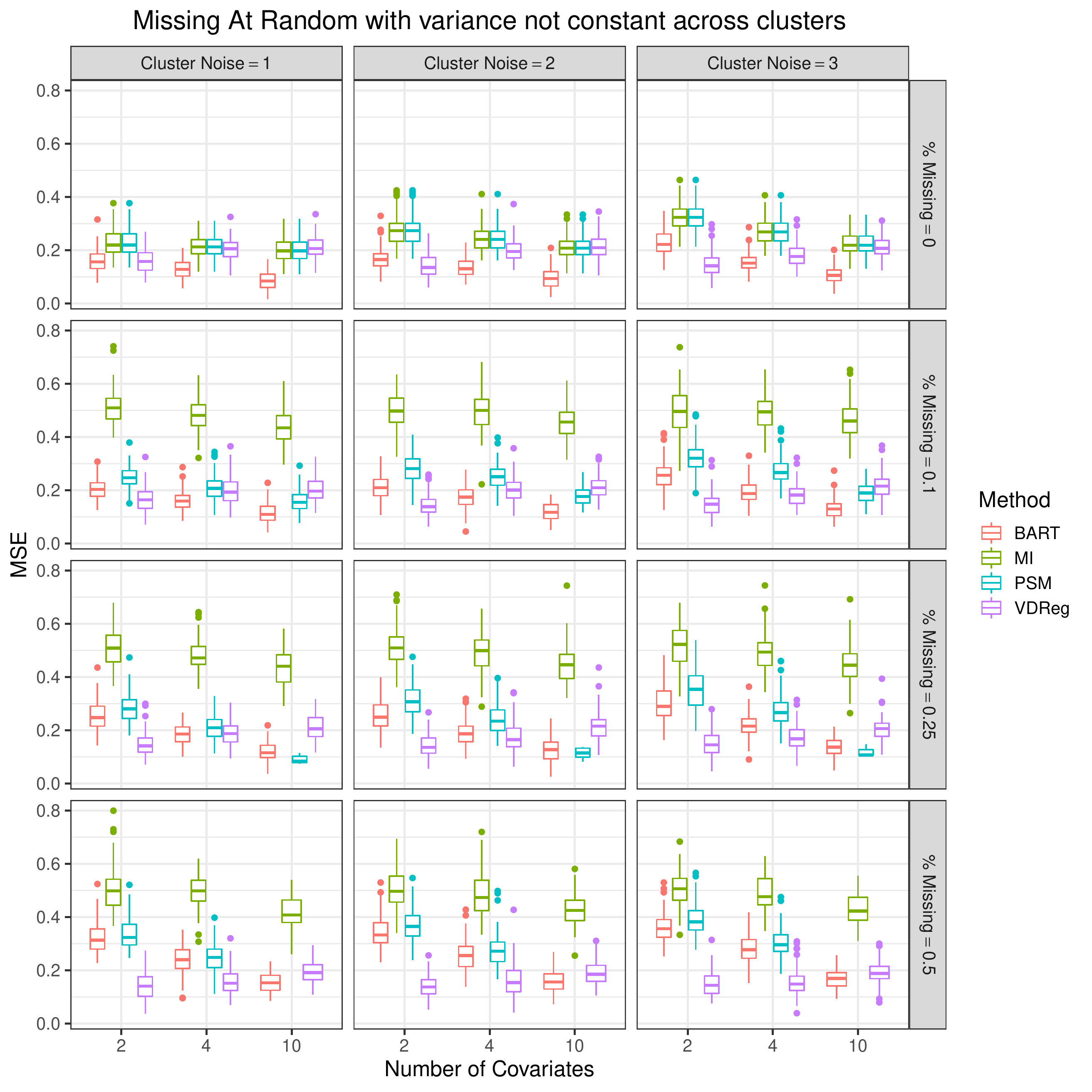}
\caption{MSE results from simulation study when missing is at random and likelihood variance ($s^2$) is {\it not} constant across the clusters.}
\label{MSE_MAR_NonConst}
\end{center}
\end{figure}

\begin{figure}[htbp]
\begin{center}
\includegraphics[scale=0.45]{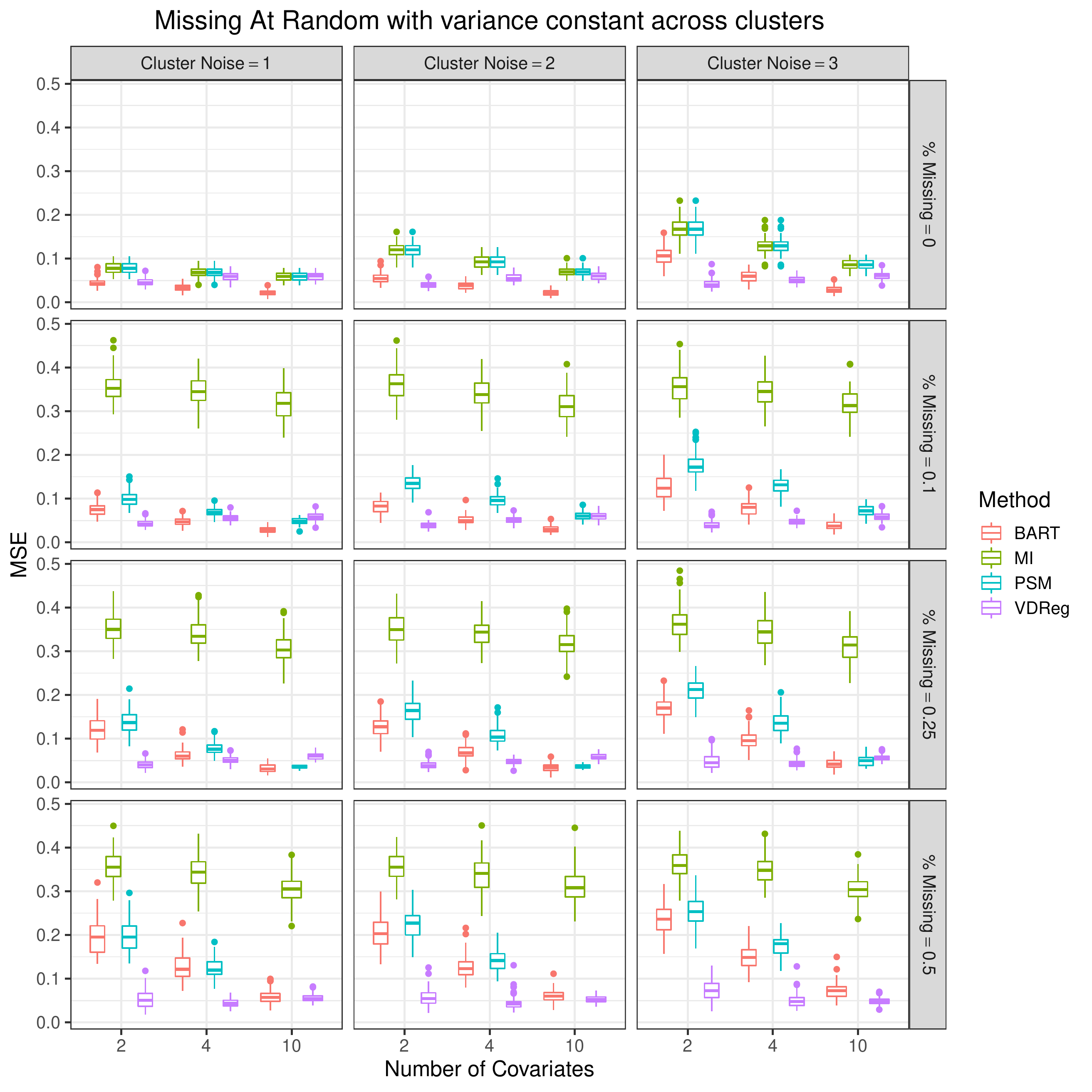}
\caption{MSE results from simulation study when missing is at random and likelihood variance ($s^2$) is  constant across the clusters.}
\label{MSE_MAR_Const}
\end{center}
\end{figure}

Next we provide figures from the simulation study that were not included
in the main document. Figures \ref{MSE_MAR_NonConst} - \ref{MSE_MAR_Const}
display results associated with MSE.  Notice that the trends described in
the main article appear in each of the figures.  Mainly, that as the
number of covariates increases, the PPMx procedure's MSE is least impacted
and the other procedures' MSE values decrease.  Also, the PPMx procedure
seems to have the lowest MSE with only two covariates while the PPMx\_PSM
has the lowest with 10 covariates and this holds regardless of the type of
missing, the percent of missingness and wether likelihood variance is
cluster-specific or not.  As expected MI (Multiple Imputation) performs
the worst across the board.

\begin{figure}[htbp]
\begin{center}
\includegraphics[scale=0.45]{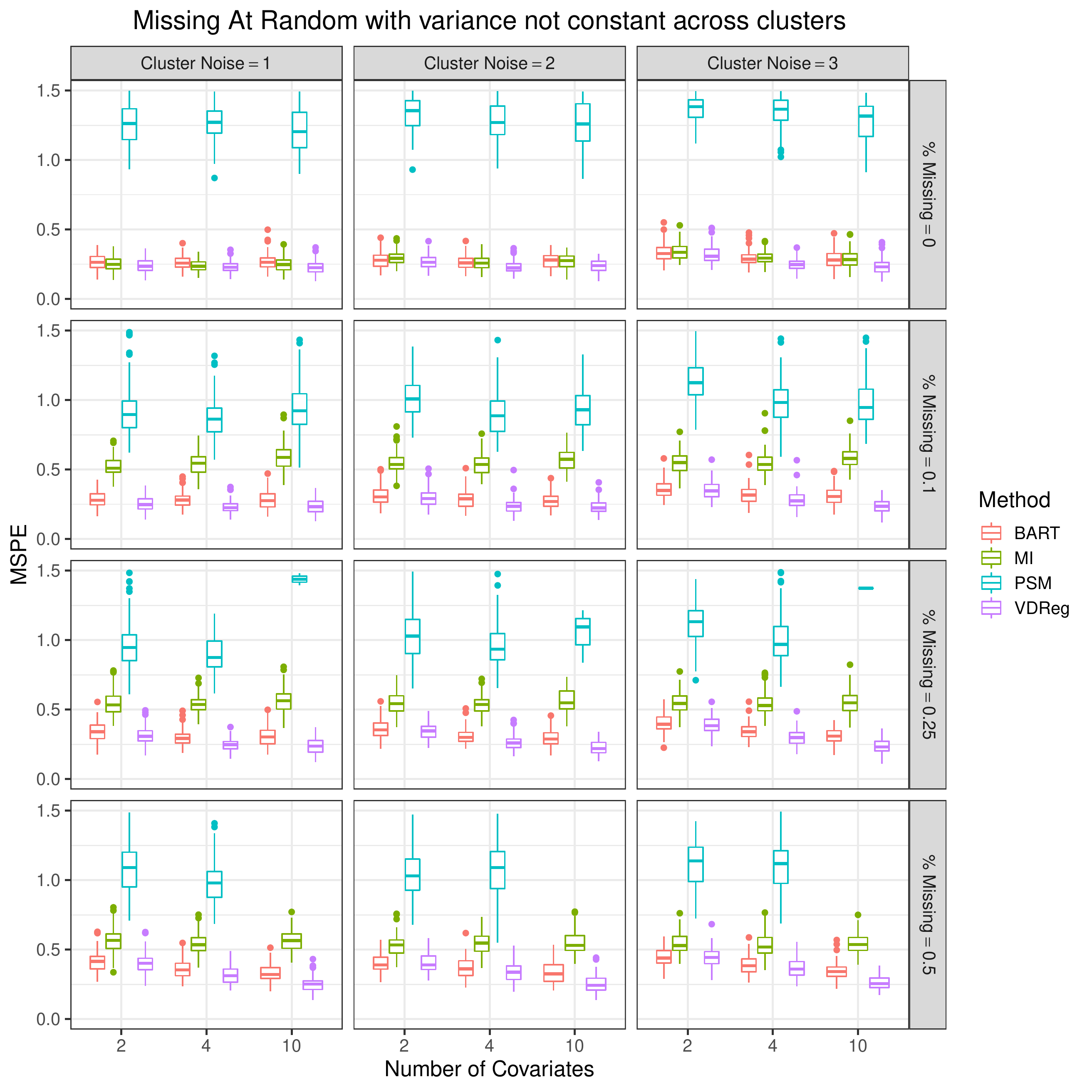}
\caption{MSPE results from simulation study when missing is at random and likelihood variance ($s^2$) is {\it not} constant across the clusters.}
\label{MSPE_MAR_NonConst}
\end{center}
\end{figure}

\begin{figure}[htbp]
\begin{center}
\includegraphics[scale=0.45]{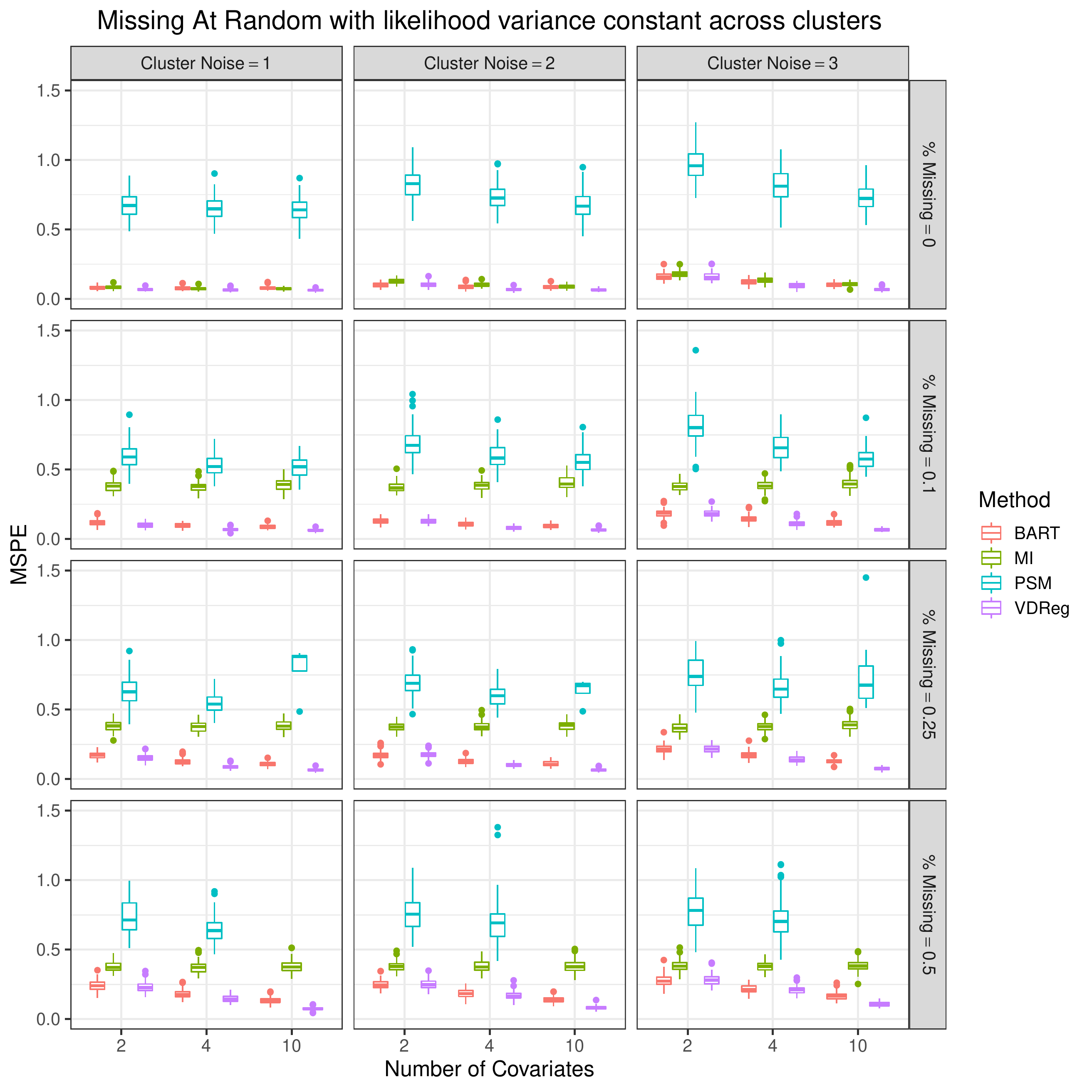}
\caption{MSPE results from simulation study when missing is at random and likelihood variance ($s^2$) is  constant across the clusters.}
\label{MSPE_MAR_Const}
\end{center}
\end{figure}

Focusing on out-of-sample prediction, Figures \ref{MSPE_MAR_NonConst} -
\ref{MSPE_MAR_Const} display results associated with the MSPE for
scenarios not included in the main article.  In terms of prediction, BART
and PPMx are the only procedures whose MSPE improves as the number of
covariates increase and this holds regardless of the type and percent of
missingness. The PPMx\_PSM and PSM procedures are negatively impacted in
terms of prediction when the number of covariates increase.  This is to be
expected because the number of missing patterns also increases as a the
number of covariates increases which results in regression coefficients
estimates being less precise and possibly biased.  Overall, it appears
that the PPMx is a very reasonable choice to make predictions in the
presence of missing data.

\begin{figure}[htbp]
\begin{center}
\includegraphics[scale=0.45]{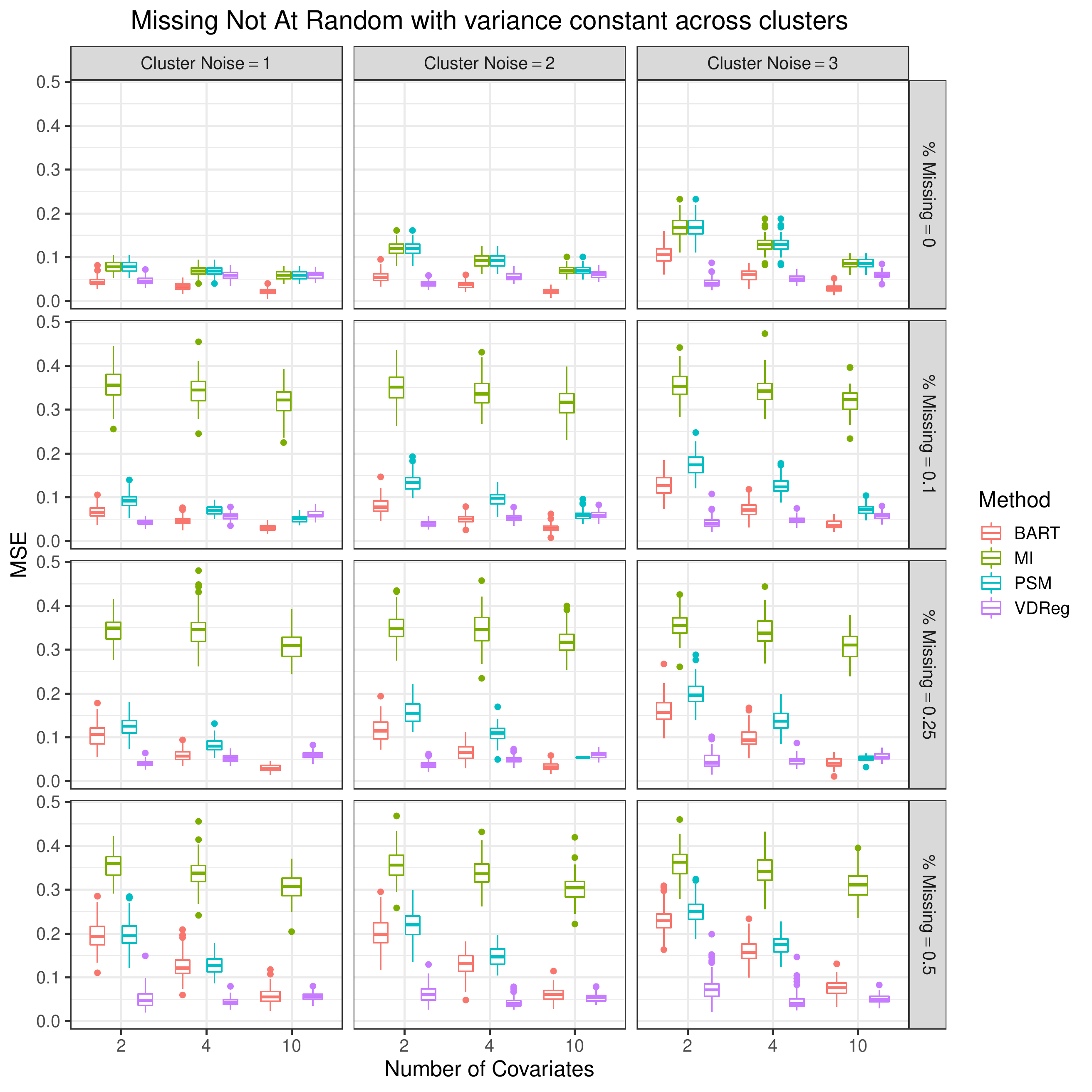}
\caption{MSE results from simulation study when missing is {\it not} at random and likelihood variance ($s^2$) is constant across the clusters}
\label{MSE_MNAR_Const}
\end{center}
\end{figure}

\begin{figure}[htbp]
\begin{center}
\includegraphics[scale=0.45]{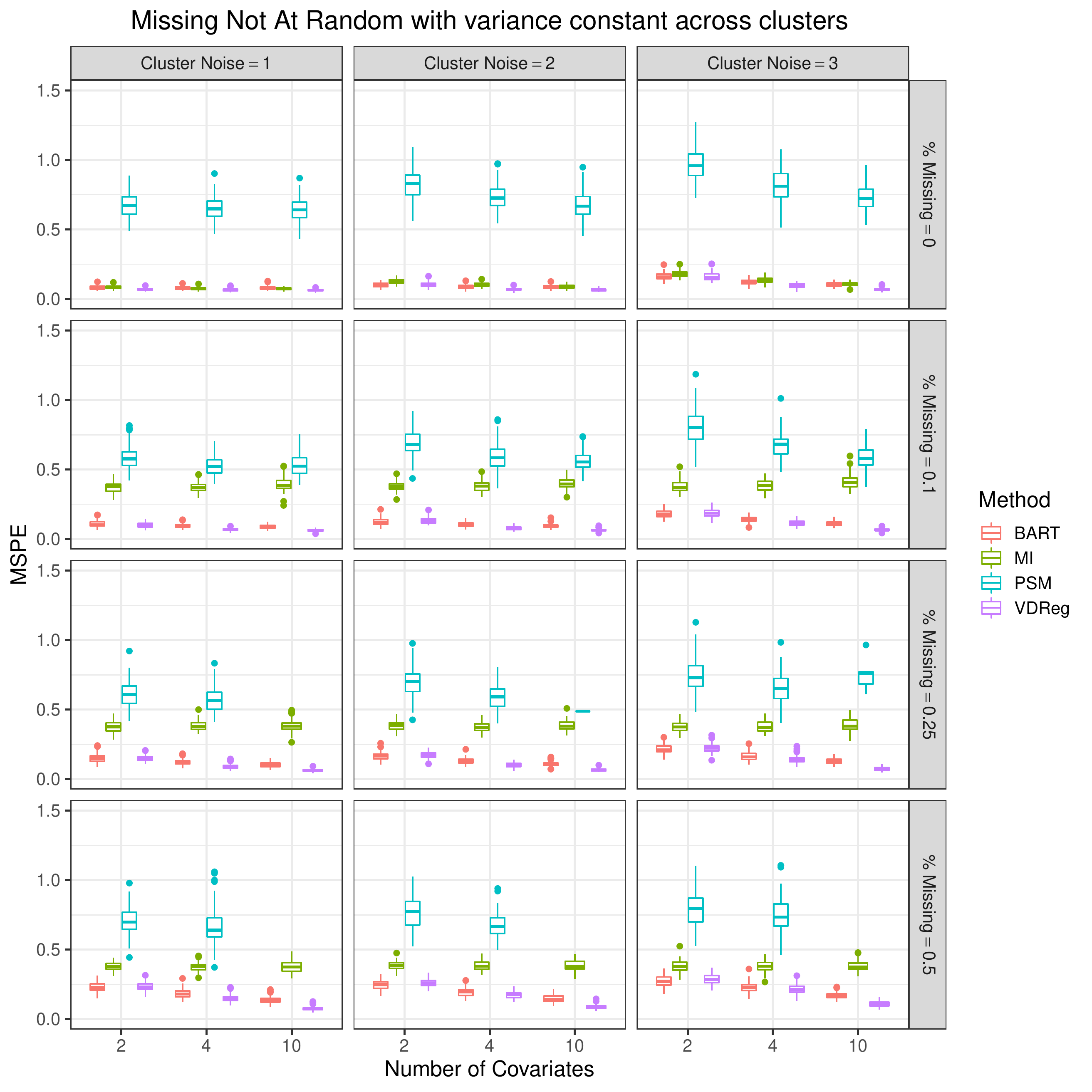}
\caption{MSPE results from simulation study when missing is {\it not} at random and likelihood variance ($s^2$) is constant across the clusters}
\label{MSPE_MNAR_Const}
\end{center}
\end{figure}

\singlespace
\bibliographystyle{plainnat}
\bibliography{reference}